\documentstyle[11pt]{article}
\newcommand{\reff}[1]{eq.~(\ref{#1})}

\newcommand{\beq}{\begin{equation}}
\newcommand{\eeq}{\end{equation}}
\newcommand{\bea}{\begin{eqnarray}}
\newcommand{\eea}{\end{eqnarray}}
\newcommand{\bq}{\begin{quote}}
\newcommand{\eq}{\end{quote}}

\renewcommand{\a}{\alpha}
\renewcommand{\b}{\beta}
\renewcommand{\c}{\gamma}

\begin{document}
\begin{titlepage}
\begin{flushright}
UQMATH-94-03\\
BI-TP-94/23\\
hep-th/9405138
\end{flushright}
\vskip.3in
\begin{center}
{\huge Solutions of the Quantum Yang-Baxter Equation with Extra
Non-Additive Parameters}
\vskip.3in
{\Large Anthony J. Bracken, Mark D. Gould, Yao-Zhong Zhang}
\vskip.1in
{\large Department of Mathematics, University of Queensland, Brisbane,
Qld 4072, Australia}
email: yzz@maths.uq.oz.au
\vskip.2in
{\Large Gustav W. Delius}
\vskip.1in
{\large Fakult\"at f\"ur Physik, Universit\"at Bielefeld, Postfach 100131,
D-33501, Bielefeld, Germany}

email: delius@physf.uni-bielefeld.de
\end{center}
\vskip.6in
\begin{center}
{\bf Abstract:}
\end{center}
We present a systematic technique to construct solutions to the
Yang-Baxter equation which depend not only on a spectral parameter but
in addition on further continuous parameters. These extra parameters
enter the Yang-Baxter equation in a similar way to the spectral parameter but
in a non-additive form.

\noindent
We exploit the fact that quantum non-compact algebras such as
$U_q(su(1,1))$ and type-I quantum superalgebras such as  $U_q(gl(1|1))$
and $U_q(gl(2|1))$ are known to admit non-trivial one-parameter
families of infinite-dimensional and finite dimensional irreps,
respectively, even for generic $q$. We develop a technique for
constructing the corresponding spectral-dependent R-matrices.
As examples we work out the the $R$-matrices for the three quantum
algebras mentioned above in certain representations.


\end{titlepage}
\newpage

\newcommand{\sect}[1]{\setcounter{equation}{0}\section{#1}}
\renewcommand{\theequation}{\thesection.\arabic{equation}}

\sect{Introduction\label{intro}}
Quantized universal enveloping algebras (quantum algebras)
\cite{Drinfeld}\cite{Jimbo}\cite{Reshetikhin} provide
a powerful tool for  finding solutions to the spectral-dependent quantum
Yang-Baxter equation (QYBE) \cite{Jimbo}\cite{BGZ90}\cite{ZGB91}.
There exists one such solution, with trigonometric dependence
on the spectral parameter, for every pair of representations of any
quantum affine Lie algebra. Through the work of many authors a large
number of such solutions has now been constructed, see references
in \cite{DGZ}. These solutions depend on a parameter $q$, the
deformation parameter of the quantum algebras. This parameter $q$
is very different in nature from the spectral parameter, because
$q$ does not enter into the Yang-Baxter equation.

It is clearly desirable to have families of solutions of the Yang-Baxter
equation depending continuously on extra parameters, entering
in a similar way to the spectral parameter. In this paper we
develop a method for the construction of such families of solutions.
The extra parameters enter the Yang-Baxter equation in a non-additive
form \footnote{The chiral Potts model has a spectral parameter which
enters the Yang-Baxter equation in a non-additive form. This
solution arises from quantum groups at $q$ a root of unity only
\cite{Bazh90}\cite{Date et al}. These are not related to our
solutions however, in which the spectral parameter stays additive
and the extra parameters are non-additive. Our solutions
exist for generic $q$.}.

Solutions of the Yang-Baxter equation have various different
applications, for example as Boltzmann weights of integrable lattice
models or as scattering matrices in integrable
quantum field theories. In all these applications the
freedom of having extra continuous parameters opens up new
and exciting possibilities. We will come back to this in the
discussion section.

The origin of the extra parameters in our solutions are the
parameters which are carried by the irreps of the associated quantum algebra.
Murakami \cite{Murakami} describes such parameters as colors carried
by irreps. For the type-I quantum superalgebras, there are nontrivial
one-parameter families of finite-dimensional irreps \cite{MS93} for generic
$q$.
For quantum simple bosonic Lie algebras families of unitary finite-dimensional
representations are possible only when the deformation parameter $q$ is
a root of unity. However, they do admit parametrized families of
unitary infinite-dimensional irreps even for generic $q$.

The main aim of this paper is to find solutions to the QYBE with extra
non-additive parameters associated with both the infinite-dimensional irreps
of a quantum simple Lie algebra and the finite-dimensional irreps of
quantum superalgebras. In section \ref{fund} we develop a
systematic and useful technique which is very much in the spirit of
the techniques in \cite{DGZ} designed to find solutions of the QYBE
acting on the tensor product module of three
different irreps of a quantum algebra. As concrete examples, we work out
the solutions ($R$-matrices) associated with a one-parameter family of
infinite dimensional irreps for the quantum non-compact algebra $U_q(su(1,1))$
in section~\ref{q-su11}, and the $R$-matrices
associated with a one-parameter family of
finite-dimensional irreps for the quantum superalgebras $U_q(gl(1|1))$
and  $U_q(gl(2|1))$ in section~\ref{q-glmn}.


\sect{General Formalism\label{fund}}
Let $G$ denote a simple Lie (super)algebra of rank $r$ with generators
$\{e_i, f_i, h_i\}$ and let $\alpha_i$ be its simple roots. Then the quantum
(super)algebra $U_q(G)$ can be defined with the structure of a
(${\bf Z}_2$-graded) quasi-triangular Hopf algebra \cite{KR89}\cite{BGZ90}.
We will not give the full
defining relations of $U_q(G)$ here but mention that $U_q(G)$ has a
coproduct structure given by
\begin{equation}
\Delta(q^{h_i/2})=q^{h_i/2}\otimes q^{h_i/2}\,,~~~\Delta(a)=a\otimes
  q^{-h_i/2}+q^{h_i/2}\otimes a\,,~~~a=e_i, f_i.
\end{equation}
The multiplication rule for the tensor product is defined for elements
$a,b,c,d\in U_q(G)$ by
\begin{equation}\label{gradprod}
(a\otimes b)(c\otimes d)=(-1)^{[b][c]}(ac\otimes bd)
\end{equation}
where $[a]\in {\bf Z}_2$ denotes the degree of the element $a$.

Let $\pi_\Phi$ be a one-parameter family of irreps of $U_q(G)$ afforded by
the irreducible module $V(\Phi)$ in such a way that the highest weight of
the irrep depends on the parameter $\Phi$.
Assume for any parameter $\Phi$ that the irrep $\pi_\Phi$
is affinizable, i.e. it can be extended to an irrep of the corresponding
quantum affine (super)algebra $U_q(\hat{G})$. Consider an operator
$R^{\Phi_1\Phi_2}(x)\in {\rm End}(V(\Phi_1)\otimes V(\Phi_2))$,
where $x\in {\bf C}$ is the usual spectral parameter and
$\pi_{\Phi_1},~\pi_{\Phi_2}$ are two irreps from the one-parameter
family. It has been
shown by Jimbo \cite{Jimbo} that a solution to the linear equations
\begin{eqnarray}
&&R^{\Phi_1\Phi_2}(x)\Delta^{\Phi_1\Phi_2}(a)=\bar{\Delta}^{\Phi_1\Phi_2}
  (a)R^{\Phi_1\Phi_2}(x)\,,~~~\forall a\in U_q(G),\nonumber\\
&&R^{\Phi_1\Phi_2}(x)\left (x\pi_{\Phi_1}(e_0)\otimes \pi_{\Phi_2}(q^{-h_0/2})+
  \pi_{\Phi_1}(q^{h_0/2})\otimes \pi_{\Phi_2}(e_0)\right )\nonumber\\
&&~~~~~~  =\left (x\pi_{\Phi_1}(e_0)\otimes \pi_{\Phi_2}(q^{h_0/2})
  +\pi_{\Phi_1}(q^{-h_0/2})\otimes \pi_{\Phi_2}(e_0)\right )R^{\Phi_1\Phi_2}
  (x)\label{r(x)1}
\end{eqnarray}
satisfies the QYBE
in the tensor product module $V(\Phi_1)\otimes V(\Phi_2)\otimes V(\Phi_3)$
of three irreps from the one-parameter family:
\begin{equation}
R^{\Phi_1\Phi_2}_{12}(x)R^{\Phi_1\Phi_3}_{13}(xy)R^{\Phi_2\Phi_3}_{23}(y)
  =R^{\Phi_2\Phi_3}_{23}(y)R^{\Phi_1\Phi_3}_{13}(xy)R^{\Phi_1\Phi_2}_{12}(x).
\end{equation}
In the above,
$\bar{\Delta}=T\cdot \Delta$, with $T$ the twist map defined by
$T(a\otimes b)=(-1)^{[a][b]}b\otimes a\,,~\forall a,b\in U_q(G)$ and
$\Delta^{\Phi_1\Phi_2}(a)=(\pi_{\Phi_1}\otimes \pi_{\Phi_2})\Delta(a)$;
also, if $R^{\Phi_1\Phi_2}(x)=\sum_i\pi_{\Phi_1}(a_i)\otimes
\pi_{\Phi_2}(b_i)$, then $R^{\Phi_1\Phi_2}_{12}(x)=\sum_i\pi_{\Phi_1}(a_i)
\otimes\pi_{\Phi_2}(b_i)\otimes I$ etc. Jimbo also showed that the solution to
(\ref{r(x)1}) is unique, up to scalar functions. The multiplicative
spectral parameter $x$ can be transformed into an additive spectral
parameter $u$ by $x=\mbox{exp}(u)$.

In all our equations we implicitly use the "graded" multiplication rule of
\reff{gradprod}. Thus the $R$-matrix of a quantum superalgebra satisfies
a "graded" Yang-Baxter equation which, when written as an ordinary
matrix equation, contains extra signs:
\begin{eqnarray}
&&\left(R^{\Phi_1\Phi_2}(x)\right)_{\a\b}^{\a'\b'}
\left(R^{\Phi_1\Phi_3}(xy)\right)_{\a'\c}^{\a''\c'}
\left(R^{\Phi_2\Phi_3}(y)\right)_{\b'\c'}^{\b''\c''}
(-1)^{[\a][\b]+[\c][\a']+[\c'][\b']}\nonumber\\
&&=\left(R^{\Phi_2\Phi_3}(y)\right)_{\b\c}^{\b'\c'}
\left(R^{\Phi_1\Phi_3}(xy)\right)_{\a\c'}^{\a'\c''}
\left(R^{\Phi_1\Phi_2}(x)\right)_{\a'\b'}^{\a''\b''}
(-1)^{[\b][\c]+[\c'][\a]+[\b'][\a']},
\end{eqnarray}
where $[\alpha]$ denotes the degree of the basis vector $v_\alpha$.
However after a redefinition
\beq\label{redef}
\left(\tilde{R}^{\Phi_1\Phi_2}\right)_{\a\b}^{\a'\b'}
=\left(R^{\Phi_1\Phi_2}\right)_{\a\b}^{\a'\b'}\,
(-1)^{[\a][\b]}
\eeq
the signs disappear from the equation. Thus any solution of the "graded"
Yang-Baxter equation arising from the R-matrix of a quantum superalgebra
provides also a solution of the standard Yang-Baxter equation after
the redefinition in \reff{redef}

Now introduce the (graded) permutation operator $P^{\Phi_1\Phi_2}$
on the tensor product
module $V(\Phi_1)\otimes V(\Phi_2)$ such that
\begin{equation}\label{gradperm}
P^{\Phi_1\Phi_2}(v_\alpha\otimes v_\beta)=(-1)^{[\alpha][\beta]}
  v_\beta\otimes v_\alpha\,,~~
  \forall v_\alpha\in V(\Phi_1)\,,~v_\beta\in V(\Phi_2)
\end{equation}
and set
\begin{equation}
\breve{R}^{\Phi_1\Phi_2}(x)=P^{\Phi_1\Phi_2}R^{\Phi_1\Phi_2}(x).
\end{equation}
Then (\ref{r(x)1}) can be rewritten as
\begin{eqnarray}
&&\breve{R}^{\Phi_1\Phi_2}(x)\Delta^{\Phi_1\Phi_2}(a)=\Delta^{\Phi_2\Phi_1}(a)
  \breve{R}^{\Phi_1\Phi_2}(x)\,,~~~\forall a\in U_q(G),\nonumber\\
&&\breve{R}^{\Phi_1\Phi_2}(x)\left (x\pi_{\Phi_1}(e_0)\otimes\pi_{\Phi_2}
  (q^{-h_0/2})+\pi_{\Phi_1}(q^{h_0/2})\otimes
  \pi_{\Phi_2}(e_0)\right )\nonumber\\
&&~~~~~~  =\left (\pi_{\Phi_2}(e_0)\otimes \pi_{\Phi_1}(q^{-h_0/2})+
  x\pi_{\Phi_2}(q^{h_0/2})\otimes \pi_{\Phi_1}(e_0)\right )
  \breve{R}^{\Phi_1\Phi_2}(x)\label{r(x)2}
\end{eqnarray}
and in terms of $\breve{R}^{\Phi_1\Phi_2}(x)$ the QYBE becomes
\begin{equation}
(I\otimes\breve{R}^{\Phi_1\Phi_2}(x))(\breve{R}^{\Phi_1\Phi_3}(xy)
  \otimes I)(I\otimes\breve{R}^{\Phi_2\Phi_3}(y))=
  (\breve{R}^{\Phi_2\Phi_3}(y)\otimes I)(I\otimes \breve{R}^{\Phi_1\Phi_3}(xy))
  (\breve{R}^{\Phi_1\Phi_2}(x)\otimes I)
\end{equation}
both sides of which act from $V(\Phi_1)\otimes V(\Phi_2)\otimes V(\Phi_3)$
to $V(\Phi_3)\otimes V(\Phi_2)\otimes V(\Phi_1)$. Note that this equation,
if written in matrix form, does not have extra signs in the superalgebra
case. This is because the definition of the graded permutation operator
in \reff{gradperm} includes the signs of \reff{redef}.

In order to
solve eqs.(\ref{r(x)2}), we use a method similar to the one developed
in \cite{DGZ} for a quite different problem. We will normalize the $R$-matrix
$\breve{R}^{\Phi_1\Phi_2}(x)$ in such a way that
\begin{equation}
\breve{R}^{\Phi_1\Phi_2}(x)\breve{R}^{\Phi_2\Phi_1}(x^{-1})=I\label{unitary}
\end{equation}
which is usually called the unitarity condition in the literature.
Consider three special cases:
$x=0,~x=\infty$ and $x=1$. For these special values of $x$,
$\breve{R}^{\Phi_1\Phi_2}(x)$ satisfies the spectral-free,
but extra non-additive parameter-dependent QYBE,
\begin{equation}
(I\otimes\breve{R}^{\Phi_1\Phi_2})(\breve{R}^{\Phi_1\Phi_3}
  \otimes I)(I\otimes\breve{R}^{\Phi_2\Phi_3})=
  (\breve{R}^{\Phi_2\Phi_3}\otimes I)(I\otimes \breve{R}^{\Phi_1\Phi_3})
  (\breve{R}^{\Phi_1\Phi_2}\otimes I).
\end{equation}
Moreover, from (\ref{r(x)2}), we have respectively, for $x=0$,
\begin{eqnarray}
&&\breve{R}^{\Phi_1\Phi_2}(0)\Delta^{\Phi_1\Phi_2}(a)=\Delta^{\Phi_2\Phi_1}(a)
  \breve{R}^{\Phi_1\Phi_2}(0)\,,~~~\forall a\in U_q(G),\nonumber\\
&&\breve{R}^{\Phi_1\Phi_2}(0)\left (
  \pi_{\Phi_1}(q^{h_0/2})\otimes\pi_{\Phi_2}(e_0)\right )
  =\left (\pi_{\Phi_2}(e_0)\otimes \pi_{\Phi_1}(q^{-h_0/2})\right )
  \breve{R}^{\Phi_1\Phi_2}(0)\label{r(0)}
\end{eqnarray}
for $x=\infty$,
\begin{eqnarray}
&&\breve{R}^{\Phi_1\Phi_2}(\infty)\Delta^{\Phi_1\Phi_2}(a)=
  \Delta^{\Phi_2\Phi_1}(a)
  \breve{R}^{\Phi_1\Phi_2}(\infty)\,,~~~\forall a\in U_q(G),\nonumber\\
&&\breve{R}^{\Phi_1\Phi_2}(\infty)\left (\pi_{\Phi_1}(e_0)\otimes\pi_{\Phi_2}
  (q^{-h_0/2})\right )=\left (\pi_{\Phi_2}(q^{h_0/2})\otimes
  \pi_{\Phi_1}(e_0)\right )
  \breve{R}^{\Phi_1\Phi_2}(\infty)\label{r(infty)}
\end{eqnarray}
and for $x=1$,
\begin{eqnarray}
&&\breve{R}^{\Phi_1\Phi_2}(1)\Delta^{\Phi_1\Phi_2}(a)=\Delta^{\Phi_2\Phi_1}(a)
  \breve{R}^{\Phi_1\Phi_2}(1)\,,~~~\forall a\in U_q(G),\nonumber\\
&&\breve{R}^{\Phi_1\Phi_2}(1)\left (\pi_{\Phi_1}(e_0)\otimes\pi_{\Phi_2}
  (q^{-h_0/2})+\pi_{\Phi_1}(q^{h_0/2})\otimes
  \pi_{\Phi_2}(e_0)\right )\nonumber\\
&&~~~~~~  =\left (\pi_{\Phi_2}(e_0)\otimes \pi_{\Phi_1}(q^{-h_0/2})+
  \pi_{\Phi_2}(q^{h_0/2})\otimes \pi_{\Phi_1}(e_0)\right )
  \breve{R}^{\Phi_1\Phi_2}(1)\label{r(1)}.
\end{eqnarray}
Eqs.(\ref{r(0)}), (\ref{r(infty)}) and (\ref{r(1)}) respectively admit a
unique solution for any given one-parameter family of irreps of
$U_q(G)$, provided such representations are consistently affinizable.
In the case of a multiplicity-free tensor product decomposition we may write
\begin{equation}
V(\Phi_1)\otimes V(\Phi_2)=\bigoplus_\mu V(\mu),
\end{equation}
where $\mu$ denotes a highest weight depending on the parameters $\Phi_1$
and $\Phi_2$.
Let $\{|e^\mu_\alpha\rangle_{\Phi_1\otimes \Phi_2}\}$ be an orthonormal
basis for $V(\mu)$ in $V(\Phi_1)\otimes V(\Phi_2)$.
$V(\mu)$ is also embedded in $V(\Phi_2)\otimes V(\Phi_1)$ through the
opposite coproduct $\bar{\Delta}$. Let
$\{|e^\mu_\alpha\rangle_{\Phi_2\otimes \Phi_1}\}$ be the
corresponding orthonormal
basis \footnote{For the precise definition of this basis see Appendix
C of \cite{DGZ}.}.
Using these bases we define operators ${\cal P}^{\Phi_1\Phi_2}_{\mu}$
and ${\bf P}^{\Phi_1\Phi_2}_{\mu}$
\begin{eqnarray}
&&{\cal P}^{\Phi_1\Phi_2}_\mu=\sum_\alpha \left |e^\mu_\alpha\right
  \rangle_{\Phi_1\otimes
  \Phi_2~\Phi_1\otimes\Phi_2}\!\left\langle e^\mu_\alpha\right |,\nonumber\\
&&{\bf P}^{\Phi_1\Phi_2}_\mu=\sum_\alpha\left |e^\mu_\alpha\right
  \rangle_{\Phi_2\otimes\Phi_1~
  \Phi_1\otimes\Phi_2}\!\left\langle e^\mu_\alpha\right |\,.
\end{eqnarray}
Clearly the ${\cal P}^{\Phi_1\Phi_2}_\mu:~V(\Phi_1)\otimes V(\Phi_2)
\rightarrow V(\mu)\subset V(\Phi_1)\otimes V(\Phi_2)$ are projection operators.
The ${\bf P}^{\Phi_1\Phi_2}_\mu:~V(\Phi_1)\otimes V(\Phi_2)\rightarrow
V(\mu)\subset V(\Phi_2)\otimes V(\Phi_1)$ are the elementary intertwiners,
i.e.,
\begin{equation}
{\bf P}^{\Phi_1\Phi_2}_\mu\Delta^{\Phi_1\Phi_2}(a)=\Delta^{\Phi_2\Phi_1}(a)
  {\bf P}^{\Phi_1\Phi_2}_\mu\,,~~~~\forall a\in U_q(G)
\end{equation}
and ${\bf P}^{\Phi_1
\Phi_2}_\mu$ and ${\cal P}^{\Phi_1\Phi_2}_\mu$ satisfy the relations
\begin{eqnarray}
&&{\bf P}^{\Phi_1\Phi_2}_\mu\,{\cal P}^{\Phi_1\Phi_2}_{\mu'}=
  {\cal P}^{\Phi_2\Phi_1}_{\mu'}\,{\bf P}^{\Phi_1\Phi_2}_\mu=\delta_{\mu\mu'}
  {\bf P}^{\Phi_1\Phi_2}_\mu\,,\nonumber\\
&&{\bf P}^{\Phi_2\Phi_1}_\mu\,{\bf P}^{\Phi_1\Phi_2}_{\mu'}=\delta_{\mu\mu'}
  {\cal P}^{\Phi_1\Phi_2}_\mu\,,\nonumber\\
&&{\cal P}^{\Phi_1\Phi_2}_\nu\,{\cal P}^{\Phi_1\Phi_2}_{\nu'}
  =\delta_{\nu\nu'}{\cal P}^{\Phi_1\Phi_2}_\nu\,,~~~~\sum_\mu
  {\cal P}^{\Phi_1\Phi_2}_\mu=I.\label{projectors}
\end{eqnarray}

We can show \cite{DGZ} that the solutions of eqs.(\ref{r(0)}),
(\ref{r(infty)}) and (\ref{r(1)}) take
the particularly simple forms,
\begin{eqnarray}
&&\breve{R}^{\Phi_1\Phi_2}(0)=\sum_{\mu}
  \epsilon(\mu)\; q^{\frac{C(\mu)-C(\Phi_1)-C(\Phi_2)}{2}}\;
  {\bf P}^{\Phi_1\Phi_2}_\mu\nonumber\\
&&\breve{R}^{\Phi_1\Phi_2}(\infty)=\sum_{\mu}
  \epsilon(\mu)\; q^{-\frac{C(\mu)-C(\Phi_1)-C(\Phi_2)}{2}}\;
  {\bf P}^{\Phi_1\Phi_2}_\mu\nonumber\\
&&\breve{R}^{\Phi_1\Phi_2}(1)=\sum_{\mu}\;
  {\bf P}^{\Phi_1\Phi_2}_\mu\label{r0-1}
\end{eqnarray}
where $C(\Lambda)=(\Lambda, \Lambda+2\rho)$ is the eigenvalue of the quadratic
Casimir invariant of $G$ in the irrep with highest weight $\Lambda$, $\rho$
is the half-sum of positive roots of $U_q(G)$, and $\epsilon(\mu)$ is the
parity of $V(\mu)$ in $V(\Phi_1)\otimes V(\Phi_2)$.

Here we illustrate the proof of the last relation in (\ref{r0-1}). From
the unitarity condition (\ref{unitary}) it follows that for $x=1$
\begin{equation}
\breve{R}^{\Phi_1\Phi_2}(1)\check{R}^{\Phi_2\Phi_1}(1)=I\,.\label{u2}
\end{equation}
We write $\breve{R}^{\Phi_1\Phi_2}(1)$ in the general form
\begin{equation}
\breve{R}^{\Phi_1\Phi_2}(1)=\sum_\mu\rho_\mu(1){\bf P}^{\Phi_1\Phi_2}_\mu
  .\label{1}
\end{equation}
Observing (\ref{u2}) we at once see that $\rho_\mu(1)$
satisfies $(\rho_\mu(1))^2=1$, so that
$\rho_\mu(1)=\pm 1$. By examining the limit $\Phi_1\rightarrow\Phi_2$,
which we can do because $\Phi_1$ and $\Phi_2$ are continuous parameters
(such arguments are not valid for the case considered in \cite{DGZ}
and thus the derivation of $\breve{R}(1)$ there is much more subtle),
and using that when $\Phi_1=\Phi_2$,~${\bf P}^{\Phi_1\Phi_1}_\mu$ are
the usual projection operators and $\breve{R}^{\Phi_1\Phi_1}(1)$ is the
identity, one can conclude that the $\rho_\mu(1)$ appearing in (\ref{1})
must equal 1 identically, thus completing the proof.

We remark that in the present case  $\Phi_1$ etc. are continuous
parameters and so
the parities $\epsilon(\mu)$ in (\ref{r0-1}) can easily be
worked out by examining the limit $\Phi_1\rightarrow\Phi_2$, in contrast
to the case considered in \cite{DGZ}.

Multiplying the second equation in (\ref{r(0)}) by ${\cal P}^{\Phi_2\Phi_1}
_\mu$ from the left and by ${\cal P}^{\Phi_1\Phi_2}_\nu$ from the right, and
using (\ref{r0-1}) and (\ref{projectors}) we obtain
\begin{eqnarray}
&&{\bf P}^{\Phi_1\Phi_2}_\mu\left (\pi_{\Phi_1}(q^{h_0/2})
  \otimes\pi_{\Phi_2}(e_0)\right ){\cal P}^{\Phi_1\Phi_2}_\mu
  ={\cal P}^{\Phi_2\Phi_1}_\mu\left (\pi_{\Phi_2}(e_0)\otimes \pi_{\Phi_1}
  (q^{-h_0/2})\right ){\bf P}^{\Phi_1\Phi_2}_\mu,\nonumber\\
&&\epsilon(\mu) q^{C(\mu)/2}{\bf P}^{\Phi_1\Phi_2}_\mu\left (
  \pi_{\Phi_1}(q^{h_0/2})\otimes\pi_{\Phi_2}(e_0)\right )
  {\cal P}^{\Phi_1\Phi_2}_\nu\nonumber\\
&&~~~~~~  =\epsilon(\nu) q^{C(\nu)/2}{\cal P}^{\Phi_2\Phi_1}_\mu
  \left (\pi_{\Phi_2}(e_0)\otimes \pi_{\Phi_1}(q^{-h_0/2})
  \right ){\bf P}^{\Phi_1\Phi_2}_\nu,~~~\forall\mu\neq\nu.\label{p-0}
\end{eqnarray}
Similarily, from (\ref{r(infty)}) and (\ref{r(1)}), we obtain
\begin{eqnarray}
&&{\bf P}^{\Phi_1\Phi_2}_\mu\left (\pi_{\Phi_1}(e_0)
  \otimes\pi_{\Phi_2}(q^{-h_0/2})\right ){\cal P}^{\Phi_1\Phi_2}_\mu
  ={\cal P}^{\Phi_2\Phi_1}_\mu\left (\pi_{\Phi_2}(q^{h_0/2})\otimes
\pi_{\Phi_1}
  (e_0)\right ){\bf P}^{\Phi_1\Phi_2}_\mu,\nonumber\\
&&\epsilon(\mu) q^{-C(\mu)/2}{\bf P}^{\Phi_1\Phi_2}_\mu\left (
  \pi_{\Phi_1}(e_0)\otimes\pi_{\Phi_2}(q^{-h_0/2})\right )
  {\cal P}^{\Phi_1\Phi_2}_\nu\nonumber\\
&&~~~~~~  =\epsilon(\nu) q^{-C(\nu)/2}{\cal P}^{\Phi_2\Phi_1}_\mu
  \left (\pi_{\Phi_2}(q^{h_0/2})\otimes \pi_{\Phi_1}(e_0)
  \right ){\bf P}^{\Phi_1\Phi_2}_\nu,~~~\forall\mu\neq\nu\label{p-infty}
\end{eqnarray}
and
\begin{eqnarray}
&&{\bf P}^{\Phi_1\Phi_2}_\mu\left (\pi_{\Phi_1}(e_0)
  \otimes\pi_{\Phi_2}(q^{-h_0/2})\right ){\cal P}^{\Phi_1\Phi_2}_\mu
  ={\cal P}^{\Phi_2\Phi_1}_\mu\left (\pi_{\Phi_2}(e_0)\otimes \pi_{\Phi_1}
  (q^{-h_0/2})\right ){\bf P}^{\Phi_1\Phi_2}_\mu,\nonumber\\
&&{\bf P}^{\Phi_1\Phi_2}_\mu\left (\pi_{\Phi_1}(e_0)
  \otimes\pi_{\Phi_2}(q^{-h_0/2})\right ){\cal P}^{\Phi_1\Phi_2}_\nu\nonumber\\
&&~~~~~~~~~~~~~  ={\cal P}^{\Phi_2\Phi_1}_\mu\left
  (\pi_{\Phi_2}(e_0)\otimes \pi_{\Phi_1}
  (q^{-h_0/2})\right ){\bf P}^{\Phi_1\Phi_2}_\nu,~~~
  \forall\mu\neq\nu.\label{p-1}
\end{eqnarray}
In deriving (\ref{p-1}), eqs.(\ref{p-0}) and eqs.(\ref{p-infty}) have been
considered.

Now the most general $\breve{R}^{\Phi_1\Phi_2}(x)$ satisfying the first
equation in (\ref{r(x)2}) may be written in the form
\begin{equation}
\breve{R}^{\Phi_1\Phi_2}(x)=\sum_{V(\mu)\in V(\Phi_1)\otimes V(\Phi_2)}
  \rho_\mu(x)\;{\bf P}^{\Phi_1\Phi_2}_\mu
\end{equation}
where $\rho_\mu(x)$, are  unknow functions depending on
$x,~q$ and the extra non-additive parameters. Inserting
the above equation into the second
equation of (\ref{r(x)2}) and multiplying the resultant equation by
${\cal P}^{\Phi_2\Phi_1}_\mu$ from the left and by
${\cal P}^{\Phi_1\Phi_2}_\nu$ from the right, and then using (\ref{p-0}),
(\ref{p-infty}) and (\ref{p-1}) to simplify the resulting equation,
one finally finds
\begin{eqnarray}
&&\left \{\rho_\mu(x)\left (xq^{C(\mu)/2}
  +\epsilon(\mu)\epsilon(\nu) q^{C(\nu)/2}\right )-
  \rho_\nu(x) \left (q^{C(\mu)/2}
  + \epsilon(\mu)\epsilon(\nu) x
  q^{C(\nu)/2}\right )\right \}\nonumber\\
&&~~~~~~~~~~\times~  {\cal P}^{\Phi_1\Phi_2}_\mu\left (
  \pi_{\Phi_1}(e_0)\otimes \pi_{\Phi_2}(q^{-h_0/2})\right )
  {\cal P}^{\Phi_1\Phi_2}_\nu=0
  \,,~~~\forall \mu\neq\nu\label{pp}
\end{eqnarray}
In many cases it is possible to determine when
${\cal P}^{\Phi_1\Phi_2}_\mu\left (
  \pi_{\Phi_1}(e_0)\otimes \pi_{\Phi_2}(q^{-h_0/2})\right )
  {\cal P}^{\Phi_1\Phi_2}_\nu\neq 0$
and thus to obtain a solution $\rho_\mu(x)$ to the system
of equations (\ref{pp}), recursively given by
\begin{equation}
\rho_\mu(x)=\rho_\nu(x)\frac{q^{C(\mu)/2}
  + \epsilon(\mu)\epsilon(\nu) xq^{C(\nu)/2} }{xq^{C(\mu)/2}
  +\epsilon(\mu)\epsilon(\nu) q^{C(\nu)/2} },~~~~\forall \mu\neq\nu.
\end{equation}
In the following sections we will work out three examples.

\sect{$U_q(su(1,1))$, Its One-Parameter Family of Irreps and
  $R$-Matrix with Non-Additive Parameter\label{q-su11}}
The quantum algebra $U_q(su(1,1))$
is defined by generators $\{e,~f,~q^{h}\}$ and relations
\begin{equation}
q^heq^{-h}=q e\,,~~q^hfq^{-h}=q^{-1}f\,,~~
[e, f]=-\frac{q^{2h}-q^{-2h}}{q-q^{-1}}
\end{equation}
with the following coproduct
\begin{equation}
\Delta(q^h)=q^h\otimes q^h\,,~~~
  \Delta(e)=e\otimes q^{-h}+q^h\otimes e\,,~~~
  \Delta(f)=f\otimes q^{-h}+q^h\otimes f
\end{equation}
Note here that we have used a slightly different normalization for the
generator $h$ compared with the one in section \ref{fund}.

Infinite-dimensional unitary irreps of $U_q(su(1,1))$ have been described
by several authors \cite{Bernard}\cite{Kalnins}\cite{Klimyk}.
The classification is similar to that in the classical
case, but there are some new features, notably the appearance of a
`strange' series of unitary irreps \cite{Klimyk} with no classical
analogue. For simplicity, we shall assume here
that $q\neq 1$ is real and positive, and consider the
infinite-dimensional unitary irreps $D^{\pm} (\Phi)$
from the so-called discrete series.
The irrep $D^{-} (\Phi)$ has a highest weight $\Phi$
but no lowest weight, and $D^{+} (\Phi)$ has a lowest
weight $-\Phi$ but no highest weight. In each case, the real
parameter $\Phi$ can take any negative value.

The structure of $D^{-} (\Phi)$ is very simple:
Let $V_{\Phi}$ be a complex Hilbert space with orthonormal basis
$\{v_{\Phi + \mu}\,;\,~\mu=0,-1,-2,\cdots\}$, and set
\begin{eqnarray}
&&hv_{\Phi +\mu}=(\Phi + \mu)v_{\Phi + \mu}\nonumber\\
&&ev_{\Phi +\mu}=([\mu]_q [\mu + 2\Phi +1]_q)^{\frac {1}{2}}
v_{\Phi +\mu + 1}\nonumber\\
&&fv_{\Phi +\mu}=([\mu -1]_q [\mu +2\Phi]_q)^{\frac {1}{2}}
v_{\Phi +\mu -1}\,.
\end{eqnarray}
where and throughout this paper,
\begin{equation}
[n]_q\equiv \frac{q^n-q^{-n}}{q-q^{-1}}\,.
\end{equation}

The tensor product $D^-(\Phi_1)\otimes D^-(\Phi_2)$ is completely reducible,
and is easily seen to be
\begin{equation}
D^-(\Phi_1)\otimes D^-(\Phi_2)=\bigoplus^{-\infty}_{\mu=0}
 D^-(\Phi_1+\Phi_2+\mu).\label{decompose1}
\end{equation}
The vector $\Omega_{\Phi_1+\Phi_2+\mu}$, corresponding
to the highest
weight $\Phi_1+\Phi_2+\mu\,,~\mu=0,-1,-2,\cdots$ in the component
$D^-(\Phi_1+\Phi_2+\mu)$, has the form
\begin{eqnarray}
\Omega_{\Phi_1+\Phi_2+\mu}&= &c \sum_{i=0}^\mu\,(-1)^i
  q^{i(\Phi_1+\Phi_2+\mu+1)}\,
  \Gamma_q^{1/2}(2\Phi_1+\mu-i+1)\,\Gamma_q^{1/2}(\mu-i)\nonumber\\
& &\cdot\Gamma_q^{1/2}(2\Phi_2+i+1)\,\Gamma_q^{1/2}(i)\,
   v_{\Phi_1+\mu-i}\otimes v_{\Phi_2+i}\,,\label{highest}
\end{eqnarray}
where $c$ is a normalization constant, and
$\Gamma _q (z)$ is a $q$-gamma function \cite{David} satisfying
$\Gamma _q (z+1) = [z]_q \Gamma _q (z)$.
To verify (\ref {highest}),
it is enough to check that $\Omega_{\Phi_1+\Phi_2+\mu}$ has
the correct weight and is annihilated by $\Delta(e)$.

Similar results hold for the structure of $D^{+}(\Phi)$ and for
the tensor product $D^+(\Phi_1)\otimes D^+(\Phi_2)$, but will not be used
here.

All irreps of $U_q(su(1,1))$ are affinizable. In particular, we have the
identification of $e_0, ~h_0$ in section \ref{fund} with $f,~h$ in this
section: $e_0\leftrightarrow f,~~h_0/2\leftrightarrow -h$. In the case  of
(\ref{decompose1}), eqs.(\ref{r0-1}) take the form
\begin{eqnarray}
&&\breve{R}^{\Phi_1\Phi_2}(0)=\sum_{\mu=0}^{-\infty}(-1)^\mu
  q^{I(\Phi_1+\Phi_2+\mu)-I(\Phi_1)-I(\Phi_2)}\;
  {\bf P}^{\Phi_1\Phi_2}_\mu\nonumber\\
&&\breve{R}^{\Phi_1\Phi_2}(\infty)=\sum_{\mu=0}^{-\infty}(-1)^\mu
  q^{-I(\Phi_1+\Phi_2+\mu)+I(\Phi_1)+I(\Phi_2)}\;
  {\bf P}^{\Phi_1\Phi_2}_\mu\nonumber\\
&&\breve{R}^{\Phi_1\Phi_2}(1)=\sum_{\mu=0}^{-\infty}\;
  {\bf P}^{\Phi_1\Phi_2}_\mu\label{r0-1'}
\end{eqnarray}
where $I(\Lambda)=\Lambda(\Lambda+1)$ is the eigenvalue of the $su(1,1)$
quadratic
Casimir invariant, $I=h(h+1)-fe=h(h-1)-ef$,
in the irrep with highest weight $\Lambda$,
and now ${\cal P}^{\Phi_1\Phi_2}_\mu:~D^-(\Phi_1)\otimes D^-(\Phi_2)\rightarrow
D^-(\Phi_1+\Phi_2+\mu)$.  Eqs.(\ref{r0-1'}) imply the identification of
$\epsilon(\mu)\,,~C(\mu)$ in (\ref{pp}):  $\epsilon(\mu)\leftrightarrow
(-1)^\mu$ and $C(\mu)/2\leftrightarrow I(\Phi_1+\Phi_2+\mu)$.

We will now determine $\rho_\mu(x)$ in (\ref{pp}) for this case.
Observe \cite{ZGB91} that the tensor
operator $\Delta(q^{-h}) (f\otimes q^h)$ behaves like a component of the
adjoint tensor operator of $U_q(su(1,1))$. Therefore,
${\cal P}^{\Phi_1\Phi_2}_\mu(\pi_{\Phi_1}(f)\otimes \pi_{\Phi_2}(q^h))
{\cal P}^{\Phi_1\Phi_2}_\nu\,,\mu\neq\nu\,,$  vanishes
unless the two highest weights
$\Phi_1+\Phi_2+\mu$ and $\Phi_1+\Phi_2+\nu$, associated with
${\cal P}^{\Phi_1\Phi_2}_\mu$ and ${\cal P}^{\Phi_1\Phi_2}_\nu$,
respectively, differ by a non-zero weight of the adjoint representation
of $U_q(su(1,1))$, that is by a root of $su(1,1)$. This implies that
${\cal P}^{\Phi_1\Phi_2}_\mu(\pi_{\Phi_1}(f)\otimes \pi_{\Phi_2}(q^h))
{\cal P}^{\Phi_1\Phi_2}_\nu\equiv 0$ for $\mu\neq\nu$
unless $\mu=\nu\pm 1$. Therefore, from (\ref{pp}),
\begin{equation}
\frac{\rho_\mu(x)}{\rho_\nu(x)}=\frac{q^{I(\Phi_1+\Phi_2+\mu)}+
  (-1)^{\mu+\nu}xq^{I(\Phi_1+\Phi_2+\nu)}}
  {xq^{I(\Phi_1+\Phi_2+\mu)}+
  (-1)^{\mu+\nu}q^{I(\Phi_1+\Phi_2+\nu)}}\,,~~~~\mu=\nu\pm 1 \,.
\end{equation}
It follows immediately that
\begin{equation}
\rho_\mu(x)=\rho_0(x)\prod_{\nu=-1}^{\mu}\frac{1-xq^{2(\Phi_1+\Phi_2+\nu)}}
{x-q^{2(\Phi_1+\Phi_2+\nu)}}\,,~~~~\mu=-1,-2,\cdots\label{solution1}
\end{equation}
which in turn leads to the quantum $R$-matrix
\begin{equation}
\breve{R}^{\Phi_1\Phi_2}(x)=\sum_{\mu=0}^{-\infty}
\prod_{\nu=-1}^{\mu}\frac{1-xq^{2(\Phi_1+\Phi_2+\nu)}}
{x-q^{2(\Phi_1+\Phi_2+\nu)}}\,{\bf P}^{\Phi_1\Phi_2}_\mu\label{final1}
\end{equation}
(it should be understood that $\prod_{\nu=-1}^0\;(\cdots)=1$,)
where the scalar factor $\rho_0(x)$ has been absorbed.

\sect{$U_q(gl(1|1)),~~U_q(gl(2|1))$, One-Parameter Families of Irreps
  and $R$-Matrix with Non-Additive Parameters\label{q-glmn}}

It is well known \cite{Kac77} that type-I superalgebras admit nontrivial
one-parameter families of finite-dimensional irreps which deform to provide
one-parameter families of finite-dimensional irreps of the corresponding
type-I quantum superalgebras \cite{MS93}. Here we are only concerned
with $U_q(gl(m|n))$, all irreps of which are known to be affinizable.

Choose $\{\varepsilon_i\}^m_{i=1}\bigcup \{\bar{\varepsilon}_j\}^n_{j=1}$ as a
basis for the dual of the Cartan subalgebra of $gl(m|n)$ satisfying
\begin{equation}
(\varepsilon_i,\varepsilon_j)=\delta_{ij},~~~~
(\bar{\varepsilon}_i,\bar{\varepsilon}_j)=-\delta_{ij},~~~~
(\varepsilon_i,\bar{\varepsilon}_j)=0\,.
\end{equation}
Using this basis, any weight $\Lambda$ may written as
\begin{equation}
\Lambda\equiv (\Lambda_1,\cdots,\Lambda_m|\bar{\Lambda}_1,
\cdots, \bar{\Lambda}_n)\equiv \sum_{i=1}^m\Lambda_i\varepsilon_i+
\sum_{j=1}^n\bar{\Lambda}_j\bar{\varepsilon}_j
\end{equation}
and the graded half sum $\rho$ of the positive roots of $gl(m|n)$ is
\begin{equation}
2\rho=\sum_{i=1}^m(m-n-2i+1)\varepsilon_i+\sum_{j=1}^n(m+n-2j+1)
\bar{\varepsilon}_j\,.
\end{equation}
In what follows we will consider the
one-parameter family of finite-dimensional irreducible
$U_q(gl(m|n))$-modules $V(\alpha)$ with highest weights of the form
$\Lambda(\alpha)=(0,\cdots,0|
\alpha,\cdots,\alpha)$.

We first consider the one-parameter family of two-dimensional irreps
$V(\alpha)$ of $U_q(gl(1|1))$ with the highest
weight $\Lambda(\alpha)=(0|\alpha)$. Assuming that $\alpha\neq 0\neq\beta,~
\alpha+\beta\neq 0$, we have the following decomposition
\begin{equation}
V(\alpha)\otimes V(\beta)=V(\Lambda_1)\bigoplus V(\Lambda_2)
\end{equation}
where $\Lambda_1=(0|\alpha+\beta)$ and $\Lambda_2=(-1|\alpha+\beta+1)$.
The Casimir operator takes the values
\begin{eqnarray}
&&C(\alpha)=-\alpha(\alpha+1),~~~C(\beta)=-\beta(\beta+1)\nonumber\\
&&C(\Lambda_1)=-(\alpha+\beta)(\alpha+\beta+1)\nonumber\\
&&C(\Lambda_2)=-(\alpha+\beta)^2-3(\alpha+\beta)\,.
\end{eqnarray}
By considering the limit $q\rightarrow 1, ~\alpha\rightarrow\beta$, it
follows that $\epsilon(\Lambda_1)=-\epsilon(\Lambda_2)=1$. Also it
is easy to conclude that ${\cal P}^{\alpha\beta}_{\Lambda_1}(\pi_\alpha(e_0)
\otimes \pi_\beta(q^{-h_0/2})){\cal P}_{\Lambda_2}^{\alpha\beta}\neq 0$
and ${\cal P}^{\alpha\beta}_{\Lambda_2}(\pi_\alpha(e_0)
\otimes \pi_\beta(q^{-h_0/2})){\cal P}_{\Lambda_1}^{\alpha\beta}\neq 0$.
Thus from (\ref{pp}) we have
\begin{equation}
\rho_{\Lambda_2}(x)=\frac{1-xq^{\alpha+\beta}}{x-q^{\alpha+\beta}}
  \rho_{\Lambda_1}(x)
\end{equation}
which gives rise to the properly normalized quantum $R$-matrix
\begin{equation}
\breve{R}^{\alpha\beta}(x)=
  {\bf P}^{\alpha\beta}_{\Lambda_1}
  +\frac{1-xq^{\alpha+\beta}}{x-q^{\alpha+\beta}}\,{\bf P}^{\alpha\beta}
  _{\Lambda_2}\,,\label{r-gl11}
\end{equation}
where again the scalar factor $\rho_{\Lambda_1}(x)$ has been absorbed.

It can be shown that the elementary intertwiners in the
above equation take the form
\begin{eqnarray}
{\bf P}^{\alpha\beta}_{\Lambda_1}&=& [\alpha+\beta]_q^{-1}\;\left (
\begin{array}{cccc}
[\alpha+\beta]_q & 0 & 0 & 0\\
0 & ([\alpha]_q[\beta]_q)^{1/2}\,q^{(\alpha+\beta)/2} & [\alpha]_q & 0 \\
0 & [\beta]_q & ([\alpha]_q[\beta]_q)^{1/2}\,q^{-(\alpha+\beta)/2} & 0 \\
0 & 0 & 0 & 0
\end{array}
\right )\,,\nonumber\\
{\bf P}^{\alpha\beta}_{\Lambda_2}&=& [\alpha+\beta]_q^{-1}\;\left (
\begin{array}{cccc}
0 & 0 & 0 & 0\\
0 & ([\alpha]_q[\beta]_q)^{1/2}\,q^{-(\alpha+\beta)/2} & -[\beta]_q & 0 \\
0 & -[\alpha]_q & ([\alpha]_q[\beta]_q)^{1/2}\,q^{(\alpha+\beta)/2} & 0 \\
0 & 0 & 0 & [\alpha+\beta]_q
\end{array}
\right )\,.\label{pp1}
\end{eqnarray}
The details of derivations and other interesting material will be published
in a separate paper \cite{BDGZ94}. With the help of (\ref{pp1}) the
R-matrix (\ref{r-gl11}) reads
\begin{equation}\label{R}
\breve{R}^{\alpha\beta}(x)=\left (
\begin{array}{cccc}
1 & 0 & 0 & 0\\
0 & -([\alpha]_q[\beta]_q)^{1/2}q^{(\alpha+\beta)/2}\cdot \frac{q-q^{-1}}
 {x-q^{\alpha+\beta}} & \frac{[\alpha]_q-[\beta]_q\left\langle \alpha+\beta
 \right\rangle}{[\alpha+\beta]_q} & 0\\
0 & \frac{[\beta]_q-[\alpha]_q\left\langle \alpha+\beta\right\rangle}
 {[\alpha+\beta]_q} &
 -([\alpha]_q[\beta]_q)^{1/2}q^{(\alpha+\beta)/2}\cdot \frac{x(q-q^{-1})}
 {x-q^{\alpha+\beta}} & 0\\
0 & 0 & 0 & \left\langle \alpha+\beta\right\rangle
\end{array}
\right )
\end{equation}
where
\begin{equation}
\left\langle\alpha+\beta\right\rangle\equiv \frac{1-xq^{\alpha+\beta}}
  {x-q^{\alpha+\beta}}\,.
\end{equation}
In the limit $x=0$, we obtain the braid group representation
$\sigma^{\alpha\beta}\equiv\breve{R}^{\alpha\beta}(0)$\,:
\begin{equation}
\sigma^{\alpha\beta}=\left (
\begin{array}{cccc}
1 & 0 & 0 & 0\\
0 & ([\alpha]_q[\beta]_q)^{1/2}q^{-(\alpha+\beta)/2}(q-q^{-1})
   & q^{-\beta} & 0\\
0 & q^{-\alpha} & 0 & 0\\
0 & 0 & 0 & -q^{-\alpha-\beta}
\end{array}
\right )
\end{equation}
For the special case $\a=\b=1$ this braid group representation has been
known. It was obtained from $U_q(gl(1|1))$ in \cite{LS91,RS92}.

We now come to the one-parameter family of four-dimensional irreps $V(\alpha)$
\cite{Links92} of $U_q(gl(2|1))$ with the highest weight $\Lambda(\alpha)
=(0,0|\alpha)$. Assuming that $\alpha\neq 0\neq\beta,~~
\alpha+\beta\neq -1$, we have the decomposition \cite{LZ}
\begin{equation}
V(\alpha)\otimes V(\beta)=V(\Lambda_1)\bigoplus V(\Lambda_2)\bigoplus
  V(\Lambda_3)
\end{equation}
where $\Lambda_1=(0,0|\alpha+\beta),~~\Lambda_2=(-1,-1|\alpha+\beta+2)$
and $\Lambda_3=(0,-1|\alpha+\beta+1)$. Thus
\begin{eqnarray}
&&C(\alpha)=-\alpha(\alpha+2),~~~C(\beta)=-\beta(\beta+2)\nonumber\\
&&C(\Lambda_1)=-(\alpha+\beta)(\alpha+\beta+2)\nonumber\\
&&C(\Lambda_2)=4-(\alpha+\beta+2)(\alpha+\beta+4)\nonumber\\
&&C(\Lambda_3)=3-(\alpha+\beta+1)(\alpha+\beta+3)\,.
\end{eqnarray}
By considering the limit $q\rightarrow 1,~~\alpha\rightarrow\beta$, we see
that $\epsilon(\Lambda_1)=\epsilon(\Lambda_2)=-\epsilon(\Lambda_3)=1$. Also
we can conclude in this case that
\begin{eqnarray}
&&{\cal P}^{\alpha\beta}_{\Lambda_1}(\pi_\alpha(e_0)\otimes
 \pi_\beta(q^{-h_0/2})){\cal P}_{\Lambda_3}^{\alpha\beta}\neq 0\nonumber\\
&&{\cal P}^{\alpha\beta}_{\Lambda_3}(\pi_\alpha(e_0)\otimes
 \pi_\beta(q^{-h_0/2})){\cal P}_{\Lambda_1}^{\alpha\beta}\neq 0\nonumber\\
&&{\cal P}^{\alpha\beta}_{\Lambda_3}(\pi_\alpha(e_0)\otimes
 \pi_\beta(q^{-h_0/2})){\cal P}_{\Lambda_2}^{\alpha\beta}\neq 0\nonumber\\
&&{\cal P}^{\alpha\beta}_{\Lambda_2}(\pi_\alpha(e_0)\otimes
 \pi_\beta(q^{-h_0/2})){\cal P}_{\Lambda_3}^{\alpha\beta}\neq 0\,.
\end{eqnarray}
Thus from (\ref{pp}) we get
\begin{eqnarray}
&&\rho_{\Lambda_3}(x)=\frac{1-xq^{\alpha+\beta}}{x-q^{\alpha+\beta}}
  \rho_{\Lambda_1}(x)\nonumber\\
&&\rho_{\Lambda_2}(x)=\frac{1-xq^{\alpha+\beta}}{x-q^{\alpha+\beta}}
  \frac{1-xq^{\alpha+\beta+2}}{x-q^{\alpha+\beta+2}}
  \rho_{\Lambda_1}(x)
\end{eqnarray}
which lead to the quantum $R$-matrix
\begin{equation}
\breve{R}^{\alpha\beta}(x)=
  {\bf P}^{\alpha\beta}_{\Lambda_1}
  +\frac{1-xq^{\alpha+\beta}}{x-q^{\alpha+\beta}}\,{\bf P}^{\alpha\beta}
  _{\Lambda_3}+
  \frac{1-xq^{\alpha+\beta}}{x-q^{\alpha+\beta}}
  \frac{1-xq^{\alpha+\beta+2}}{x-q^{\alpha+\beta+2}}\,{\bf P}^{\alpha\beta}
  _{\Lambda_2}
\end{equation}
where again the scalar factor $\rho_{\Lambda_1}(x)$ has been absorbed.
We have explicit expressions for the elementary intertwiners
appearing in this expression as $16\times 16$ matrices which we will
publish in \cite{BDGZ94}.

\sect{Conclusion\label{concl}}

We have developed a systematic technique for constructing solutions
to the QYBE with extra non-additive parameters. The technique is a
generalization of that used in \cite{DGZ} for a different problem.
We have treated in particular solutions associated to families of
representations of the quantum algebras $U_q(su(1,1)),~U_q(gl(1|1))$
and $U_q(gl(2|1))$.
The technique can be applied to other quantum
(super)algebras with families of irreducible highest weight
or lowest weight representations. We can also apply the technique
to families of representations of Yangians and will then obtain
rational $R$-matrices with extra parameters.
Our general expressions (\ref{r0-1}) for the braid generators can also be
used to construct multivariable link invariants from quantum (super)algebras.

We expect that the possibility of extra parameters in solutions of the
Yang-Baxter equation will open up many new applications. The physical
interpretation of the extra parameters will depend on the particular
application. An example:

The scattering matrices for quantum excitations in integrable 2-dimensional
quantum field theories are given by solutions of the Yang-Baxter
equation. The ratios between the masses of the quantum excitations are
determined by the locations of the poles of the S-matrices. Because
these poles are fixed in all known non-trivial crossing symmetric
$R$-matrices, these $R$-matrices are able to describe only theories in which
the
ratios of the quantum masses are fixed to particular values. In our
R-matrices the locations of the poles depend on the extra parameters.
So if our R-matrices are used as S-matrices, these parameters have
the interpretation of adjustable quantum masses in integrable
theories. We intend to study this possibility in detail.

\vskip.3in
\begin{center}
{\bf Acknowledgements:}
\end{center}
Y.Z.Z. thanks Jon R. Links and Rui Bin Zhang for discussions. The financial
support from the Australian Research Council is gratefully acknowledged.

\newpage

\end{document}